\documentclass[9pt,twocolumn,twoside]{pnas-new}

\usepackage[capitalise]{cleveref}
\newcommand{\gv}[1]{\ensuremath{\mbox{\boldmath$ #1 $}}} 
\newcommand{\bv}[1]{\ensuremath{\boldsymbol{#1}}} 

\newcommand{\rom}[1]{\uppercase\expandafter{\romannumeral #1\relax}}

\usepackage{soul}

\templatetype{pnasresearcharticle} 

\title{Multi-curvature viscous streaming: flow topology and particle manipulation}

\author[a,1]{Yashraj Bhosale}
\author[a,1]{Giridar Vishwanathan}
\author[a]{Tejaswin Parthasarathy} 
\author[a,2]{Gabriel Juarez} 
\author[a,b,c,2]{Mattia Gazzola}

\affil[a]{Mechanical Science and Engineering, University of Illinois at Urbana-Champaign, Urbana, IL 61801, USA}
\affil[b]{National Center for Supercomputing Applications, University of Illinois at Urbana-Champaign, Urbana, IL 61801, USA}
\affil[c]{Carl R. Woese Institute for Genomic Biology, University of Illinois at Urbana-Champaign, Urbana, IL 61801, USA}

\leadauthor{Bhosale, Vishwanathan} 

\significancestatement{
Viscous streaming, an inertial phenomenon, refers to the steady flows that emerge when a fluid oscillates around a localized microfeature. There, stresses are concentrated, remodeling the surrounding flow over short time/length scales. The result is an efficient flow machinery to mix, transport, trap, or sort at the microscale. Nonetheless, classically, viscous streaming has been investigated for uniform-curvature bodies, for which only two distinct flow topologies exist, constraining its utility and scope. We combine predictive simulations and versatile experimental techniques to demonstrate how the rational design of multi-curvature bodies unlocks a vast range of flow topologies, demonstrated here in compact and tunable microparticle filtration and separation devices. Overall, our technology transcends previous flow/particle manipulation limitations, significantly expanding viscous streaming potential.

}

\authorcontributions{Y.B., T.P., and M.G. designed the computational research; G.V. and G.J. designed the experimental research; Y.B. performed numerical computations and analyzed data; G.V. performed experimental research and analyzed data; Y.B., G.V., T.P., G.J., and M.G. wrote the paper}
\authordeclaration{The authors declare no competing interests.}
\equalauthors{\textsuperscript{1}Y.B. and G.V. contributed equally to this work.}
\correspondingauthor{\textsuperscript{2}To whom correspondence should be addressed. E-mail: mgazzola@illinois.edu, gjuarez@illinois.edu}


\begin{abstract}

Viscous streaming refers to the rectified, steady flows that emerge when a liquid oscillates around an immersed microfeature, typically a solid body or a bubble. The ability of such features to locally concentrate stresses produces strong inertial effects to which both fluid and immersed particles respond within short length ($\mathcal{O}(100) ~ \mathrm{\mu}$m) and time ($\mathcal{O}(10^{-3})$ s) scales, rendering viscous streaming arguably the most efficient mechanism to exploit inertia at the microscale. Despite this potential, viscous streaming has been investigated in rather narrow conditions, mostly making use of bodies of uniform curvature (cylinders, spheres) for which induced flow topologies are constrained to only two regimes, classically referred to as single and double layer regimes. This severely limits the scope of potential applications, and sits in stark contrast to inertial focusing (the approach that perhaps has most defined the field of inertial microfluidics), where instead boundaries with multiple curvatures have long been utilised to magnify inertial effects, expand design space, and enable novel applications. Here, we demonstrate that a multi-curvature approach in viscous streaming dramatically extends the range of accessible flow topologies. We show that numerically predicted, but never experimentally observed, streaming flows can be physically reproduced, computationally engineered, and in turn used to enhance particle manipulation, filtering and separation in compact, robust, tunable and inexpensive devices. Overall, this study provides an avenue to unlock the utility of viscous streaming with potential applications in manufacturing, environment, health and medicine, from particle self-assembly to microplastics removal.

\end{abstract}

\dates{This manuscript was compiled on \today}
\doi{\url{www.pnas.org/cgi/doi/10.1073/pnas.XXXXXXXXXX}}

\begin{document}

\maketitle
\thispagestyle{firststyle}
\ifthenelse{\boolean{shortarticle}}{\ifthenelse{\boolean{singlecolumn}}{\abscontentformatted}{\abscontent}}{}


\dropcap{T}he recognition that, at the microscale, small but finite inertia can be employed to manipulate flows and suspended particles, has majorly impacted microfluidics \cite{di2007continuous, di2009inertial, martel2014inertial, zhang2016fundamentals}. 
Inertial effects have been central to the development of purely passive, hydrodynamic control strategies to align \cite{ozkumur2013inertial}, separate \cite{wang2011size}, concentrate \cite{ookawara2010process} and mix \cite{lutz2003microfluidics} liquids, particles and chemicals, delivering high throughputs, operational simplicity, cost effectiveness, precision and delicacy (biological samples) in a robust, fault-tolerant fashion. Because of these appealing features, modern inertial microfluidics has found  broad application across disciplines, from engineering \cite{seo2007membrane} and biology \cite{masaeli2012continuous} to environment \cite{schaap2012lab}, health and medicine \cite{mach2010continuous}.

Two main flow phenomena may be leveraged to harness inertia at the microscale: inertial focusing and viscous streaming. Inertial focusing relies on cross-streamline particle migration (from the competition between wall lift and shear gradient forces) and microchannel curvature variations to focus particles of given properties at key downstream locations, from which they are collected and processed \cite{di2007continuous, martel2014inertial}.
Due to its success, particularly in clinical and point-of-care settings \cite{di2007continuous,martel2014inertial}, inertial focusing has perhaps come to define the field of inertial microfluidics. Nonetheless, limitations remain: indeed, inertial focusing operates over relatively large length ($\sim \mathcal{O}(10)$ cm) and time ($\sim \mathcal{O}(1)$ s) scales \cite{di2009inertial, martel2014inertial}; it typically relies on highly specialized microchannels, rendering it rather inflexible to new target particles and applications; it is poorly tunable, with the flow rate as the sole real-time control parameter, limiting opportunities for modulation, rapid and localized on/off activations, and modular integration. Finally, its design process is time consuming and mostly reliant on trial-and-error, due to the current paucity of  reliable, accurate and predictive simulations \cite{bazaz2020computational}.

Potential solutions may be instead offered by viscous streaming, a complementary and perhaps overlooked inertial flow phenomenon. Viscous streaming refers to the steady, rectified flows that emerge when a fluid oscillates around a localised microfeature, typically a solid body or a bubble \cite{longuet1998viscous, lutz2005microscopic, wang2012efficient}. Such microfeatures have the ability to concentrate stresses \cite{marmottant2003controlled, marmottant2004bubble}, thus distort and remodel, even dramatically, the surrounding flow and its topology \cite{parthasarathy2019streaming,bhosale2020shape,chan2021three}. The result is a remarkably consistent, controllable and convenient machinery to shape both flow and particle paths. Indeed, since the flow reorganises over short length-scales (microfeature size), a spectrum of high velocity curvatures and gradients is produced, which in turn generate strong inertial forces on particles \cite{agarwal2021unrecognized}. This allows for precise, selective manipulation over compact footprints ($\mathcal{O}(100) ~ \mathrm{\mu}$m) at the millisecond scale ($\mathcal{O}(10^{-3}) ~$s) \cite{thameem2016particle,thameem2017fast}, making streaming arguably the most efficient way to exploit inertia at the microscale.

Despite its potential, viscous streaming has been narrowly explored, with applications \cite{lutz2003microfluidics,lutz2005microscopic,lutz2006characterizing} overwhelmingly employing classically understood bodies of uniform curvature (cylinders, spheres) which admit only two resulting flow topologies, known as the single and double layer regimes \cite{holtsmark1954boundary,lane1955acoustical}. Such a limited reportoire restricts the scope of achievable microfluidic applications, and sits in stark contrast to inertial focusing, where instead the importance of boundaries (channel walls) varying in curvature has long been recognised and leveraged to expand application range and improve performance \cite{di2009inertial,martel2014inertial,bazaz2020computational}.

Only recently, computational studies have begun to explore the nexus between body-curvature manipulation and streaming flow reorganization, revealing a rich variety of flow topologies and dynamics \cite{parthasarathy2019streaming,bhosale2020shape,chan2021three}, never experimentally reported and of potential practical use. Here, we lay out a technology based on localised microfeature design, directly driven flow oscillations, and independently controlled frequencies and amplitudes, to experimentally access such numerically predicted flows. 
After demonstrating their existence, we further showcase their utility in computationally engineered filtering and separation devices that are tunable, versatile and effective. 
Overall, this study fundamentally advances our understanding of viscous streaming and paves the way for its expansion in practical settings, while contributing to bridging the gap between experimental and computational inertial microfluidics.                  

\begin{figure*}
\centering
\includegraphics[width=\linewidth]{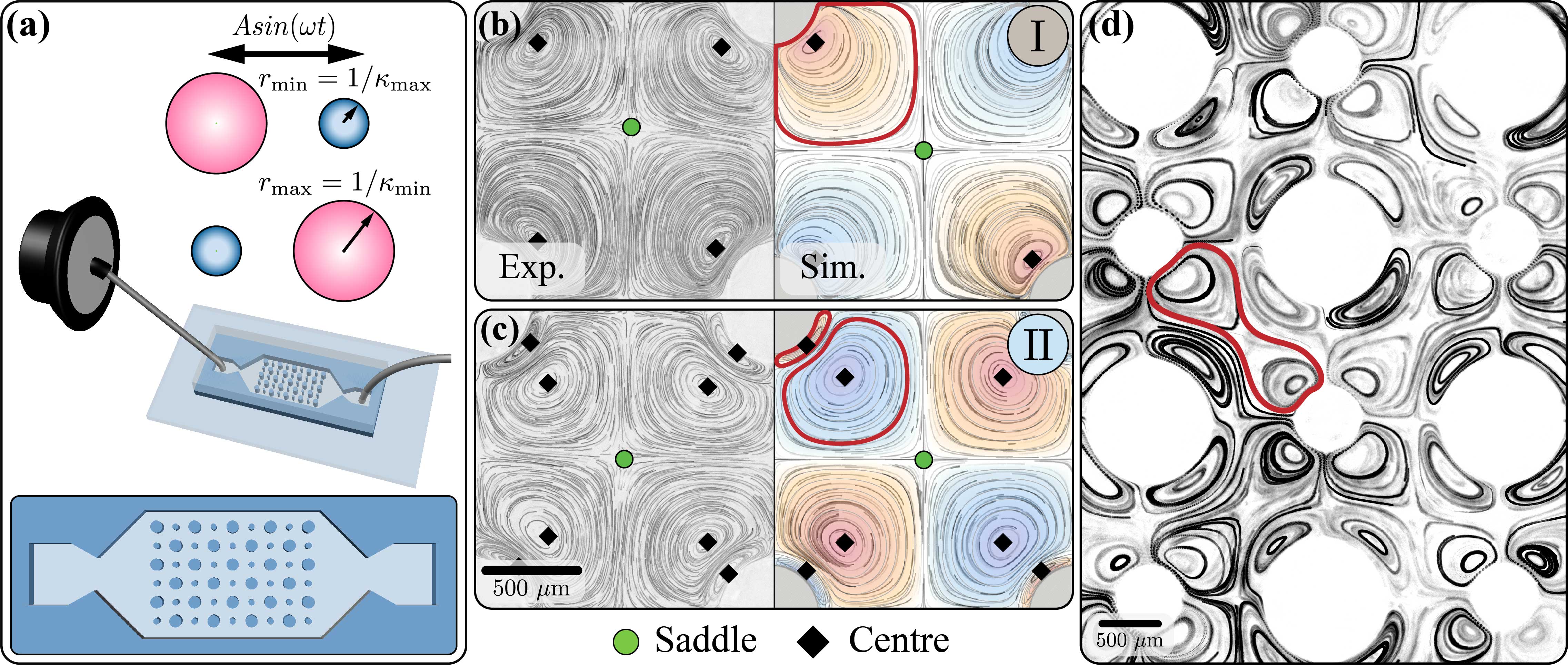}
\caption{
Streaming flows in lattice arrays. 
(a) Illustration of the lattice geometry that consists of alternating cylinders. 
The repeating unit cell entails cylinders of two different radii, therefore curvatures $\kappa_{max}$ and $\kappa_{min}$, and a fixed centre-to-centre spacing. 
The experimental setup is constituted of a glass-bonded PDMS channel connected to a loudspeaker that generates oscillatory flows along the horizontal direction. 
(b, c) Time-averaged particle pathlines observed in experiments (left) and simulations (right), for a constant curvature lattice system with $\kappa_{max} / \kappa_{min} = 1$. 
Two distinct flow topologies, direct generalization of the well-known (b) single layer (referred to as Phase \rom{1} later) and (c) double layer (referred to as Phase \rom{2} later) regimes are reported.
The flow topologies are four-fold symmetric, and defining saddles (green circles) and centres (black diamonds) are labeled. Blue represents clockwise rotation, while orange stands for counter-clockwise rotation. 
(d) Time-averaged particle pathlines observed in experiments across multiple unit cells for a lattice system with $\kappa_{max} / \kappa_{min} = 2$.
}
\label{fig:setup}
\end{figure*}

\section*{Results}

\subsection*{Multiple discrete curvatures -- The lattice system} Towards exploring the effects of multiple body-curvatures on viscous streaming, a simplified setup referred to as the lattice system is considered \cite{bhosale2020shape}. The lattice system consists of a 2D array of circular cylinders, with exactly two distinct curvatures $\kappa_{max}$ and $\kappa_{min}$, arranged in a checkerboard pattern, and immersed in a background oscillatory flow of amplitude $A$ and angular frequency $\omega$, as shown in \cref{fig:setup}(a). The center-to-center distance between adjacent cylinders is kept constant at $6.25  / \kappa_{max}$ throughout the study, without loss of generality \cite{bhosale2020shape}. This setup allows the injection of multiple, discrete body-curvatures in the flow system, whose response, at small amplitudes ($A\ll1 / \kappa_{max}$), is completely specified by only two parameters: the non-dimensional Stokes layer thickness $\delta_{AC} \kappa_{max} = \kappa_{max} \sqrt{\nu / \omega}$, where $\nu$ is the fluid kinematic viscosity, and the curvature ratio $\kappa_{max} / \kappa_{min}$. Thus, by keeping $\kappa_{max}$ fixed while modifying $\kappa_{min}$ and $\omega$, the effects of curvature variations ($\kappa_{max} / \kappa_{min}$) and flow conditions ($\delta_{AC} \kappa_{max}$) on streaming topology can be studied in a controlled, systematic manner. A rich phase space composed of hitherto unseen streaming flow topologies was recently discovered numerically, and underlying bifurcation mechanisms understood via dynamical systems theory \cite{bhosale2020shape}. To examine the potential for microfluidic applications, a realization of the lattice system is attempted in this study.

\subsubsection*{Experimental realization} To systematically access multi-curvature viscous streaming regimes, an external driver is used to generate oscillatory flows within a PDMS microfluidic channel \cite{Vishwanathan2020}, sculpted with a lattice structure at its center. A loudspeaker is directly connected to a microfluidic tubing, the other end of which is inserted into the device, as shown in \cref{fig:setup}(a). The vibration of the loudspeaker diaphragm generates a time-varying pressure that produces harmonic displacement of the liquid within the device, at independently controlled frequencies and amplitudes.

Three lattice systems are fabricated, consisting of rectangular arrays of $10\times 8$ cylinders with $r_{min}=\kappa_{max}^{-1} = 287 \mathrm{\ \mu m}$ and $r_{max}=\kappa_{min}^{-1} = 287,\ 333,\ 571 \mathrm{\ \mu m}$, resulting in the curvature ratios $\kappa_{max}/\kappa_{min} \approx 1.0$, $1.2$, and $2.0$. The investigated range of flow conditions $\delta_{AC}\kappa_{max}$ is achieved with oscillation frequencies of $50 - 600$ Hz, and deionized water as the working liquid. For all cases, the oscillation amplitude ($A\leq 50\ \mu$m) is small compared to the cylinder radii. 

\begin{figure*}
\centering
\includegraphics[width=\linewidth]{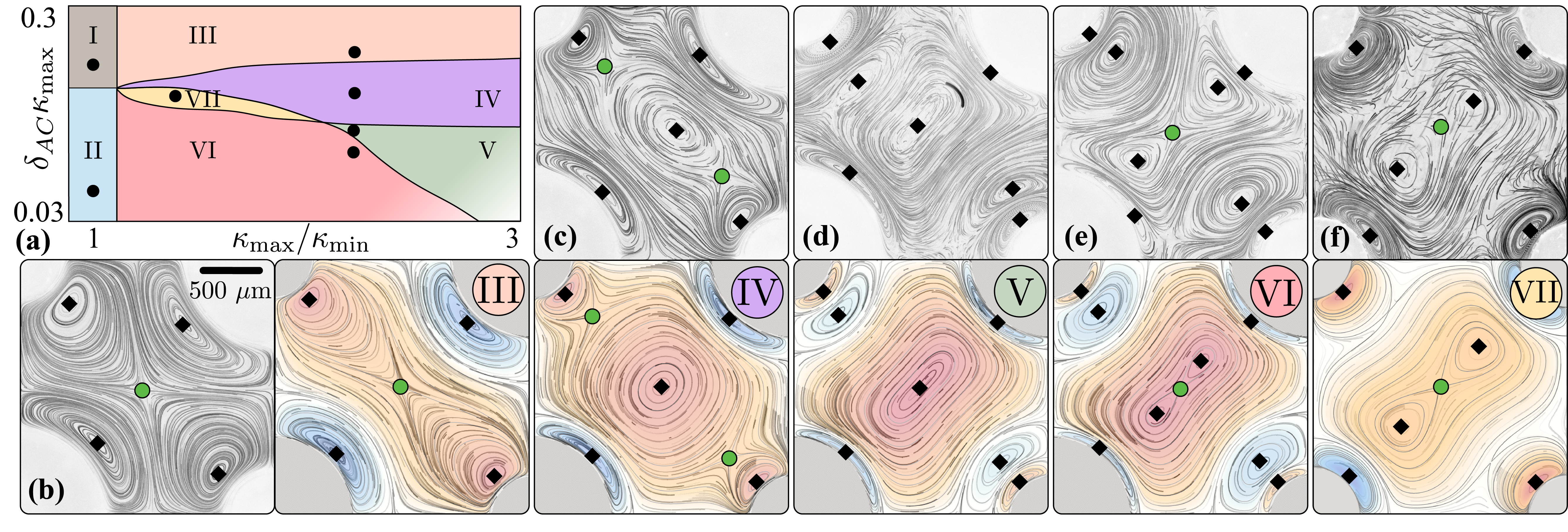}
\caption{
Phase space of streaming flow topologies, identified through the lattice system. 
(a) Phase space as a function of $\delta_{AC} \kappa_{max}$ and $\kappa_{max} / \kappa_{min}$. 
Black lines indicate transition boundaries between phases, and black dots indicate the specific point in the phase space reported here. Streaming flows are classified into distinct phases depending on their flow topology, characterised by saddles (green circles) and centres (black diamonds). (b-f) Time-averaged particle pathlines observed in experiments (left) and simulations (right) for distinct phases, indicated by roman numerals. Phases in (b,c,d,e) are obtained for $\kappa_{max}/\kappa_{min}=2$, while (f) employs $\kappa_{max}/\kappa_{min}=1.2$. Topological markers (saddles and centres) are identified and indicate good agreement between experiments and simulations.
}
\label{fig:ps}
\end{figure*}

\subsubsection*{Phase space} We start with the intuitive case of the uniform-curvature lattice ($\kappa_{max}/\kappa_{min}$=1), illustrated in \cref{fig:setup} (b,c). On the left, observed streaming flow pathlines for a unit cell are shown, while on the right, corresponding numerical streamlines are juxtaposed, with blue and orange representing clockwise and counter-clockwise rotations, respectively. Of particular relevance are the flow critical points, where the velocity is zero. In two-dimensional incompressible flows, saddles (green circles) and centers (black diamonds) conveniently offer a sparse yet complete representation of the flow field, its topology and underlying dynamics. Further, and importantly, they are of practical use with regards to mixing, trapping or transport. Specifically, centers are employed to attract and retain particles \cite{chong2013inertial,chong2016transport}, while saddles (and connecting streamlines) partition the flow, enabling particle separation \cite{lutz2003microfluidics,thameem2016particle} or targeted mixing in spatially controlled chemistry \cite{lutz2003microfluidics,lutz2006characterizing}.

We first observe that the cases of \cref{fig:setup} (b) and (c), corresponding to $\delta_{AC} \kappa_{max} = 0.19$ and $\delta_{AC} \kappa_{max}=0.06$, both exhibit four-fold symmetry, in line with the symmetry of the unit cell. At relatively large $\delta_{AC} \kappa_{max}$(>0.1), streaming flows result in single vortices (orange, outlined in red) with distinct centers, neatly separated by saddles. This topology is an expected, direct generalisation of the single-layer regime for individual cylinders \cite{holtsmark1954boundary}. At smaller $\delta_{AC} \kappa_{max}$, additional counter-rotating outer vortices (blue, outlined in red) appear diagonally, squeezing the original vortices into inner bounded regions adjacent to the cylinders, as shown in \cref{fig:setup}c. This is a generalisation of the double-layer regime for individual cylinders \cite{holtsmark1954boundary}. As can be noticed, experiments accurately replicate numerical solutions.

After establishing our approach in the uniform scenario, lattice systems with multiple curvatures are explored. As illustrated over six unit cells in \cref{fig:setup}(d), a curvature ratio departure from unity ($\kappa_{max}/\kappa_{min}=2$) breaks four-fold symmetry, leading to a two-fold diagonal symmetry instead. Evidently, reduced symmetry modifies flow topology, permitting the transport of material across two of the four vortices. This is a significant breakthrough from the limitations of the uniform curvature lattice, and provides an avenue to usefully employ multiple body-curvatures for sculpting, manipulating and connecting flow regions.

Systematic variation of flow conditions ($\delta_{AC} \kappa_{max}$) and curvature ratio ($\kappa_{max} / \kappa_{min}\geq$ 1) yield the computationally-determined phase space of \cref{fig:ps}(a), in which seven distinct flow topologies are identified. Here, we experimentally probe their existence (black dots), spanning curvature ratio by means of our three lattice channels ($\kappa_{max} / \kappa_{min} = $ 1, 1.2 and 2). For a given lattice, different flow topologies (phases) can be accessed by modifying $\delta_{AC}  / \kappa_{max}$ via the frequency $\omega$, with low frequencies corresponding to large $\delta_{AC}\kappa_{max}$, and vice-versa.
The case of $\kappa_{max} / \kappa_{min} = 1$ has already been discussed, and we refer to the topologies of \cref{fig:setup}(b) and (c) as Phase \rom{1} and \rom{2}, respectively. Of the remaining five phases, Phases \rom{3}-\rom{6} are achieved with the lattice of $\kappa_{max} / \kappa_{min} = 2$, while Phase \rom{7} is achieved with $\kappa_{max} / \kappa_{min} = 1.2$. 

Phase \rom{3} is observed for $\delta_{AC}  / \kappa_{max} > 0.2$. In this phase, the inner vortices of the smaller cylinders interact with each other and form a closed connected bicentric region (marked in orange with two centers and a saddle), as shown in \cref{fig:ps}(b). This bicentric region then separates the inner vortices (blue) of the larger cylinders.

Phase \rom{4} is observed when $0.09 < \delta_{AC} \kappa_{max}< 0.2$. Here, a single central vortex (orange with a center) flanked by two saddles (green) is observed, as shown in \cref{fig:ps}(c). This central vortex then separates the inner vortices (blue) of the larger cylinders.

Phase \rom{5} is observed when $0.07 < \delta_{AC} \kappa_{max}< 0.09$. In this case, a central vortex (orange with a center) identical to the one in Phase \rom{4} is present, as shown in \cref{fig:ps}(d). Additionally, new outer vortices (blue with centers) appear on either side of the central one, in the vicinity of the smaller cylinders.

Phase \rom{6} is seen for $\delta_{AC} \kappa_{max} < 0.07$. In this phase, a bicentric recirculation zone (orange, two centers and a saddle) is observed, as shown in \cref{fig:ps}(e). Additionally, outer vortices (blue with centers) are seen on either side of this recirculation zone, around the smaller cylinders.

Phase \rom{7} is realized for $\delta_{AC} \kappa_{max} \approx 0.11$ and $\kappa_{max}/\kappa_{min}=1.2$, and shown in \cref{fig:ps}(f). Once again, a bicentric recirculation zone (orange, two centers and a saddle) is seen, but unlike Phase \rom{6}, no outer vortices are present near the smaller cylinders.

These streaming flows, never experimentally observed before, are found to well-agree with numerical predictions \cite{bhosale2020shape}, thus establishing their viability and robustness for potential applications. Indeed, they allow to expose suspended particles to a sequence of highly varied flow environments, which in turn can be leveraged (particularly for inertial particles, Stokes number > 0.3) to induce, hasten or modulate cross-streamline migration for manipulation purposes. In purely oscillatory flows, and for sufficiently long time scales, all inertial particles are eventually trapped at vortex centers, a fact previously exploited for secure positioning \cite{lutz2006hydrodynamic}. In the presence of commonly employed finite transport flows, however, particles' residence time in the device is limited and trapping may or may not occur depending on initial location, and relative strength and topology of the streaming flow. This implies that appropriate streaming patterns, perhaps enabled by our lattice system, may be better suited than others to selectively and effectively remove incoming particles from the mean flow, thus acting as contactless, compact filters, which we present next.

\begin{figure*}[t!]
\centering
    \includegraphics[width=\linewidth]{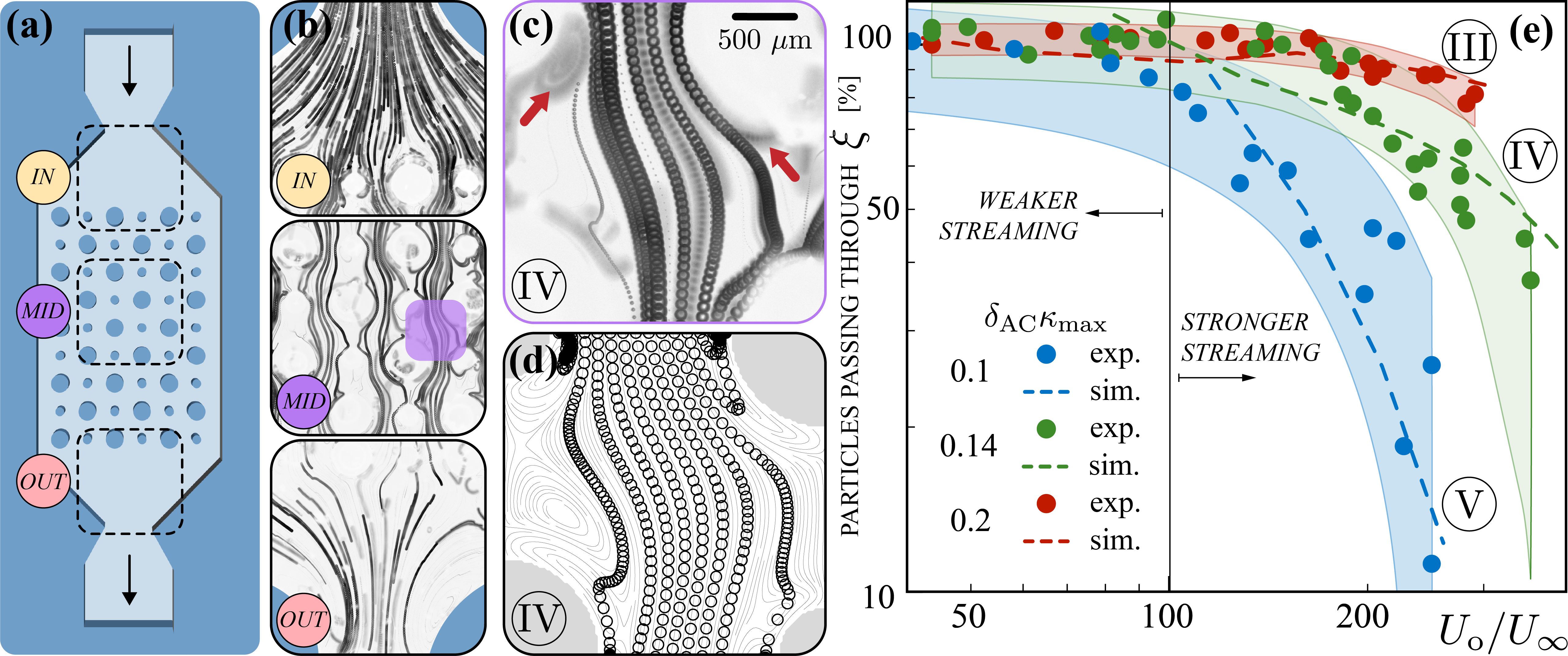}
\caption{
Particle filtration with streaming flows in lattice arrays. 
(a) Illustration of the lattice array used and the locations examined for particle concentration. (b) Particle pathlines observed in experiments within the device, as the particle suspension flows from inlet to outlet. The reduction in pathlines density, from inlet to outlet, indicates particle trapping within the device. Example particle trajectories in (c) experiments and (d) simulations for $\kappa_{max}/\kappa_{min}=2$,  $\delta_{AC}\kappa_{max}=0.14$ (Phase \rom{4}), and $U_o/U_{\infty}=200$. Particle trapping locations are indicated by red arrows. (e) Filtration efficiency $\xi= c_{\textrm{out}} / c_{\textrm{in}}$ measured experimentally (dots), corresponding envelopes (shaded regions, \( 95\)\% confidence intervals of a second-order polynomial fit to the experimental data), and predicted simulation slopes (dashed lines, with an intercept shift of $\sim 20\%$ along both axes) as a function of $U_o/U_{\infty}$ (oscillations vs. transport flow velocities) for three different lattice phases (\rom{3}, \rom{4} and \rom{5}). Particle fluxes across control surfaces at the inlet ($c_{in}$) and outlet ($c_{out}$) are evaluated using standard particle tracking velocimetry. The effect of flow topology (lattice phase) manifests as markedly different filtration behaviours.
}
\label{fig:filter}
\end{figure*}

\subsubsection*{Particle filtration}
To experimentally examine the operation of the lattice system as a filtration device, the previously described $\kappa_{max} / \kappa_{min} = 2$ lattice channel is divided into three major sections of observation, labelled as \textit{IN}, \textit{MID} and \textit{OUT} (\cref{fig:filter}a). The first zone, \textit{IN}, refers to the inlet of the microchannel, through which a neutrally buoyant suspension of 65 $\mu$m polystyrene particles in 22\% w/w glycerol/water solution is injected at a volumetric flow rate of 0.1 ml min$^{-1}$. These inertial particles are tracked over a period of 50 s, and the input particle flux $c_{\textrm{in}}$ across a fixed cross-section is evaluated (details in figure caption). The second zone, \textit{MID}, refers to the core region of the lattice system where significant trapping occurs. Finally, the third zone, \textit{OUT}, refers to the outlet of the filter, where the output particle flux $c_{\textrm{out}}$ is evaluated similarly to the input flux. Filtration efficiency $\xi = c_{\textrm{out}} / c_{\textrm{in}}$ is then defined as the relative number of particles flowing through the system, where low values of $\xi$ indicate high filtration rates, and vice-versa.

Without oscillatory flow, virtually all particles pass through the lattice and $\xi\sim 100\%$. When instead the filter is activated by turning the loudspeaker on, complex time-averaged flow topologies emerge from the interplay between transport flow and streaming, leading to particle retention within few milliseconds (\cref{fig:filter}b). Indeed, as can be appreciated in \cref{fig:filter}(c), inertial particles are extracted from the mean flow, and indefinitely captured within the streaming vortices (red arrows), a mechanism confirmed in simulations (\cref{fig:filter}d).

Filtering efficiency is then systematically characterised via experiments and simulations as a function of $U_o/U_{\infty}$, which expresses the relative strengths between streaming ($U_o = A \omega$, oscillation velocity) and transport flow ($U_{\infty}$, free-stream velocity), and $\delta_{AC} \kappa_{max}$, which determines the streaming flow topology. 

Three values of $\delta_{AC} \kappa_{max}$ are chosen, corresponding to Phases \rom{3}, \rom{4} and \rom{5}, and for each, $U_o/U_{\infty}$ is varied from 50 to 400. Corresponding filtering efficiencies $\xi$ are plotted in \cref{fig:filter}(e), where dots are experimental measurements, and shaded regions are their envelopes (caption for details). As expected, for relatively weak streaming ($U_o/U_{\infty} < 100$), nearly all particles pass through the filter for any phase considered. When instead the streaming strength is increased, the system response is dramatically altered. For Phase \rom{3} (red), filtration rate improves slowly, remaining relatively ineffective with over $\xi > 80\%$ of the particles flowing through. For Phases \rom{4} (green) and \rom{5} (blue) instead, filtration rates improve markedly, with about $35\%$ and $10\%$ of the particles passing through, respectively. When compared with simulation results (dashed lines), we find qualitative agreement, with the efficiency slopes correctly captured. Finally, we note that performance can be further improved by simply adding a few rows of cylinders.

It is worth appreciating that by varying $\delta_{AC} \kappa_{max}$ by a factor of 2  at $U_o/U_{\infty}=200$, an increase in filtration of nearly two orders of magnitude is numerically predicted and indeed experimentally attained. 
This step-change in filtration performance reflects a change in flow topology, specifically from the less effective Phase \rom{3} to the more effective Phases \rom{4} and \rom{5}. We attribute the efficacy of the latter phases to the presence of a central vortex (orange, \cref{fig:ps} c-d), otherwise absent in Phase \rom{3}. The mean flow reconfigures this vortex, which is pushed aside strengthening the bottom/top vortices near the cylinders, as illustrated in \cref{fig:filter} (d). This, in turn, facilitates trapping at the indicated locations (red arrows, \cref{fig:filter}).

The extraordinary sensitivity afforded by transitioning across streaming topologies, combined with their ability to almost instantaneously ($< 10$ ms) reconfigure upon frequency $\omega$ variations, can be harnessed to dynamically control particle distributions in time and space in a purely hydrodynamic fashion. This makes streaming an adaptive and flexible mechanism, uniquely complementary to current inertial microfluidic techniques.

Having established the utility of multi-curvature streaming via the case of discrete curvature configurations, we proceed to demonstrate a distinct microfluidic application, based on a continuous spectrum of curvatures, at the single-object level.

\begin{figure*}[ht!]
\centering
\includegraphics[width=1\linewidth]{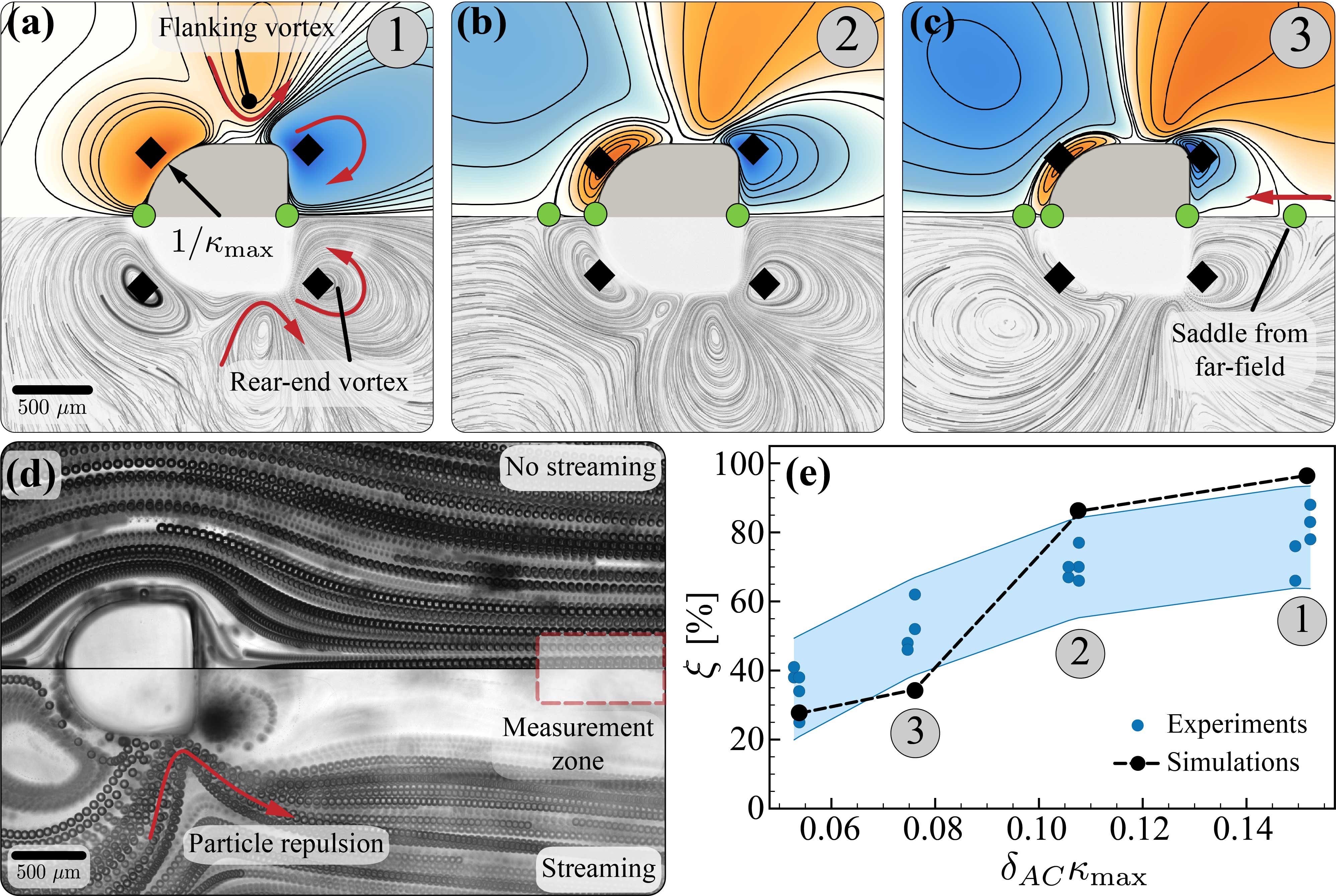}
\caption{
    Particle separation using the bullet system. (a-c) Comparison of simulated and experimental streaming pathlines around a bullet shaped cylinder, for Phases \textbf{1}, \textbf{2} and \textbf{3}, respectively. (d) Demonstration of inertial separation of particles from the wake region at $\delta_{AC} \kappa_{max} = 0.05$. (e) Efficiency of particle separation $\xi$, defined as the number of particles that flow in the measurement zone, located in the wake of the bullet, relative to the unperturbed (i.e. no streaming) reference particle content, plotted against $\delta_{AC} \kappa_{max}$ for $U_o/U_{\infty}=100$. Data points represent experimental results, the shaded region provides the corresponding envelope (\( 95\)\% confidence intervals of a second-order polynomial fit to the experimental data), and dashed lines show simulation results.
}
\label{fig:separation}
\end{figure*}

\subsection*{Continuous curvature variation -- The bullet system}
The bullet refers to the object formed by the extrusion of a single convex shape created by joining a semi-circle of diameter $2/\kappa_{max}$ to a rounded semi-square of same side-length, as illustrated in \cref{fig:separation} (a). This particular geometry was rationally designed in-silico to generate streaming features conducive to enhanced particle transport \cite{parthasarathy2019streaming,bhosale2020shape}. Here the bullet system is experimentally realized and subsequently employed for continuous particle separation, an unattainable application in the case of purely circular geometries.

\subsubsection*{Experimental realization}
We employ the same oscillatory flow generation and channel microfabrication techniques of the lattice. The bullet is characterized by a major radius of $1/\kappa_{max}=0.5$ mm, smoothing radius at the corners of 0.1 mm, and extrusion height of 2.5 mm.

\subsubsection*{Phase space}
Analogous to the lattice system, as $\delta_{AC} \kappa_{max}$ is decreased, distinct flow topologies emerge. We refer to them as Phase \textbf{1}, \textbf{2} and \textbf{3}, as illustrated in \Cref{fig:separation}(a-c). We emphasize the remarkable agreement between computationally designed (top half) and experimentally realised (bottom half) flow fields.

At higher values of $\delta_{AC} \kappa_{max}$, Phase \textbf{1} is observed. In this phase, a single set of vortices (center, black diamonds), which resemble the inner vortices observed for a circular cylinder, are seen on the circular front-end, as shown in \cref{fig:separation}(a). On the square rear-end, due to the large curvature mismatch at the corners, a set of vortices of large size and flow strength is obtained. In addition to these rear-end vortices, a new set of outer vortices flank the bullet sides. Decreasing $\delta_{AC} \kappa_{max}$ results in Phase \textbf{2}. In this phase (\cref{fig:separation}b), the original set of vortices at the circular front-end reconfigures as inner bounded vortices (thickness defined by green saddles), surrounded by outer counter-rotating vortices, similar to the double-layer regime of a circular cylinder. Additionally, a well-defined boundary appears between inner vortices on the circular front-end and flanking vortices. By further decreasing $\delta_{AC} \kappa_{max}$, the system transitions to Phase \textbf{3} (\cref{fig:separation}c), where the square-end vortices become bounded by the saddle approaching from the far-field. Consequently, flanking vortices increase in strength and curvature near the rear-end. 

The bullet illustrates how rapid computational prototyping enables new and specialised sets of flow topologies, via microfeature design. We anticipate that, upon application of a left-to-right transport flow, the flanking vortices will be effective in laterally deflecting incoming particles. Additionally, the rear vortices are expected to trap remaining undeflected particles, rendering the bullet well-suited for continuous separation, which we present next. 

\subsubsection*{Particle separation}
To experimentally demonstrate particle separation, a steady transport flow laden with $65\ \mu$m neutrally buoyant, inertial particles is generated past the bullet, such that the direction of flow is from the circular-end towards the square-end. Particle separation is then activated by superimposing streaming onto the mean flow. This results in the example flow topology of \cref{fig:separation} (d). As particles are convected around the bullet, they encounter the flanking vortices, which, as predicted, are observed to deflect particles aside. Any remaining particle that makes it to the square-end of the bullet is then readily trapped by the rear vortices. This creates a wake region mostly free from particles, referred to as the separation zone. We emphasize the contrast between streaming application and lack thereof, illustrated in \cref{fig:separation}(d).

Quantification of particle separation is performed similarly to the lattice system, where particles are now tracked in a measurement zone (marked in  \cref{fig:separation}d) before and after streaming is applied. Here, separation efficiency $\xi$ refers to the number of particles that flow in the measurement zone relative to the unperturbed (i.e. no streaming) reference particle content, and depends once again on $U_o/U_{\infty}$ and $\delta_{AC} \kappa_{max}$.  For the fixed value $U_o/U_{\infty} = 100$, the efficiency $\xi$ is plotted as a function of $\delta_{AC} \kappa_{max}$ in \cref{fig:separation}(e). As it can be observed, drastically different behaviors emerge across phases, with Phase \textbf{3} being evidently superior ($< 30\%$ of the original particles found in the measurement zone), relative to the much more modest performance of Phase \textbf{1} ($\sim 90 \%$ of the particles still present in the measurement zone). We attribute this difference to the bounded and stronger rear-end vortices of Phase \textbf{3}, accompanied by stronger flanking vortices. Finally, we note the quantitative (within error margins) agreement with simulations.

\section*{Conclusion}

This study fundamentally advances our understanding of viscous streaming beyond classical settings, through a combination of predictive simulations and versatile experimental techniques. It illuminates the role of multiple body-curvatures in establishing and manipulating unconventional flow topologies, dramatically expanding the intrinsically limited repertoire afforded by traditional uniform-curvature streaming. It illustrates the utility of injecting ranges of curvature in the flow, both in a discrete (lattice) and continuous (bullet) sense, at the level of multiple or individual objects, to achieve contactless, purely hydrodynamic, and real-time tunable microparticle filtration and separation. It demonstrates an inexpensive, robust and simply controllable technological implementation. It establishes the viability and quantitative accuracy of in-silico design, enabling rapid computational prototyping. All in all, these innovations provide the tools and understanding to expand the use of viscous streaming beyond its current niche set of applications, thus realising its full potential, characterized by vigorous inertial effects that can be rapidly activated, spatially and temporally controlled, and modularly composed, synergistically with mainstream inertial microfluidics.

\matmethods{
\subsection*{Simulation method and numerical implementation}
We briefly recap the governing equations and the numerical solution technique. We consider incompressible viscous flows in a periodic or unbounded domain $\Sigma$. In this fluid domain, immersed solid bodies perform simple harmonic oscillations. The bodies are density matched and have support $\Omega$ and boundary $\partial \Omega$ respectively. The flow can then be described 
using the incompressible Navier–Stokes equations \cref{eqn:ns}
\begin{equation}
\bv{\nabla} \cdot \gv{u} = 0 ; ~~ \frac{\partial \gv{u}}{\partial t}+(\gv{u} \cdot \bv{\nabla}) \gv{u} = -\frac{\bv{\nabla} P}{\rho} + \nu \nabla^{2} \gv{u}, ~~ \gv{x} \in \Sigma \backslash \Omega \label{eqn:ns}
\end{equation}
where $\rho$, $P$, $\gv{u}$ and $\nu$ are the fluid density, pressure, velocity and kinematic viscosity, respectively. The dynamics of the fluid–solid system is coupled via the no-slip boundary condition $\gv{u} = \gv{u_s}$, where $\gv{u_s}$ is the solid body velocity. The system of equations is then solved using a velocity–vorticity formulation with a combination of remeshed vortex methods and Brinkmann penalization \cite{gazzola2011simulations}. This method has been validated across a range of flow–structure interaction problems, from flow past bluff bodies to biological swimming \cite{gazzola2011simulations,gazzola2012flow,gazzola2012c,gazzola2014reinforcement,gazzola2016learning,bhosale2021remeshed}. Recently, the accuracy of this method has been demonstrated in rectified flow contexts as well, capturing steady streaming responses from arbitrary rigid shapes in 2D and 3D \cite{parthasarathy2019streaming,bhosale2020shape,chan2021three}.
Lastly, the motion of inertial particles demonstrated in the lattice trapping and bullet separation sections, is captured using the Maxey-Riley equations \cite{maxey1983equation}. 

\subsection*{Experimental Methods}
The channels were fabricated using Polydimethylsiloxane (Sylgard 184\texttrademark, PDMS, 10:1 resin:crosslinker) molded from a CNC micromachined aluminium mold, pretreated with silicone mold release spray.
The PDMS was cleaned with isopropanol, perforated, and then bonded to a glass slide after oxygen plasma treatment for 2 minutes. 
Polyethylene tubing (PE 200\texttrademark, ID 1.5 mm, OD 1.9 mm) was inserted at the inlet and outlet.

A solution of 22\% (w/w) glycerol in deionized water with a density and kinematic viscosity of 
of $\rho=1.15 \ \mathrm{kg/m^3}$ and $\nu=1.68 \ \mathrm{mm^2/s}$, respectively, was used. 
Polystyrene particles were suspended and remained neutrally buoyant in the aqueous solution.
Particles with a diameter of $5 \ \mathrm{\mu m}$ were used as tracers since the particle response time $\tau = \rho d^2 / 18 \mu$ was equal to $1.6$ $\mu$s, much less compared to the oscillatory timescales considered in this study.
Particles with a diameter of $65 \ \mathrm{\mu m}$ were used for particle manipulation experiments since the particle response time was equal to $0.4$ ms. 

Sinusoidal oscillatory flow of frequencies ranging from $50-600$ Hz was produced at the outlet by an external oscillatory driver generated by a loudspeaker diaphragm. 
Details of the generation and fidelity of oscillatory flow are reported elsewhere \cite{Vishwanathan2020}.  
Brightfield microscopy with $2 \times$ and $4 \times$ objective lens (depth of field $125 \mathrm{\ \mu m}$) were used to image tracer particles at the channel mid plane. 
Frame rates between 1-20 fps were used with an exposure time of $100 \ \mathrm{\mu}$s to minimize streaking.

}

\showmatmethods{}

\acknow{
We thank S. Hilgenfeldt for helpful discussions over the course of this work.
The authors acknowledge support by the National Science Foundation under NSF CAREER Grant No. CBET-1846752 (MG) and by the Blue Waters project (OCI- 0725070, ACI- 1238993), a joint effort of the University of Illinois at Urbana-Champaign and its National Center for Supercomputing Applications. 
This work used the Extreme Science and Engineering Discovery Environment (XSEDE) \cite{towns2014xsede} Stampede2, supported by National Science Foundation grant no. ACI-1548562, at the Texas Advanced Computing Center (TACC) through allocation TG-MCB190004.
}

\showacknow{}

\bibliography{pnas-sample}

\end{document}